\documentclass[11pt,doublespacing]{article}
\usepackage{geometry}                		
\pdfoutput=1

\usepackage{graphicx}				
\usepackage{amssymb}
\usepackage[square,authoryear]{natbib}
\usepackage[titletoc,title]{appendix}
\usepackage{lineno}

\title{Comment on `Heating of Enceladus due to the dissipation of ocean tides' by R. Tyler}
\author{Mikael Beuthe\\
\it Royal Observatory of Belgium,\\
\it Avenue Circulaire 3, 1180 Brussels, Belgium\\
\it E-mail: mikael.beuthe@observatoire.be}      
\date{\small Published in Icarus 350 (2020) 113934. doi:10.1016/j.icarus.2020.113934}					

\begin{document}
\maketitle

\begin{abstract}
Dissipation of ocean potential energy is proposed by \citet{tyler2020} as a new mechanism leading to possible high-power states for Enceladus.
I show here that this process actually results from viscoelastic dissipation within the crust.
For plausible values of Enceladus's ocean thickness, crustal dissipation can be computed with the standard approach of static deformations of solid layers by equilibrium tides.
\end{abstract}


\section{Introduction}

I will comment on several claims made in `Heating of Enceladus due to the dissipation of ocean tides' \citep{tyler2020}, which I summarize with the following extracts:
\begin{enumerate}
\item `The ability to associate dissipation with the ocean tidal potential energy extends the previous work (...) to form two basic families under which generic tidal solutions can be explored.'
\item `Tyler follows the precedent in terrestrial oceanography and considers ``ocean tidal dissipation'' to refer to the dissipation of ocean tidal energy -- regardless of the ultimate location where this energy is deposited or converted to heat.'
\item `Even if the dissipation takes place mostly in the overlying ice layer, it is predominantly or solely the motion of the ocean driving that of the ice'
\item `in the viscoelastic membrane ice model, for example, the ice is massless and therefore it is not forced directly by the tidal forces'
\item `The results show plausible states whereby the ocean tides (even if Enceladus is covered with thick ice) generate the observed heat flux.'
\end{enumerate}
I start by giving general counterarguments to the above points, before delving into the technical justification.
\begin{enumerate}
\item Dissipation of ocean potential energy is not a new dissipation mechanism but is in fact equal to crustal dissipation in a viscoelastic shell.
The most obvious argument is that this kind of dissipation directly depends on the rheology of the viscoelastic crust (through the imaginary part of the effective shear modulus of the crust), and thus vanishes if the crust is elastic.
This argument is supported by a complete computation of crustal dissipation as the product of stress and strain rate in the shell.
Since stress and strain are each proportional to the radial displacement of the bottom of the shell, crustal dissipation is proportional to the squared radial displacement of the top of the ocean. This explains why crustal dissipation looks a bit like dissipation of ocean potential energy.
\item In the planetology literature, with the exception of Tyler's papers, ocean dissipation always refers to dissipation associated with fluid motion in the ocean.
It is important to note that dissipation of ocean potential energy still occurs in the static limit of equilibrium tides, for which only dissipation within solid layers is significant.
Not only is it confusing to call `ocean dissipation'  heat that is dissipated inside solid layers, but it gives the impression that one should compute it with a dynamical approach (for example the Laplace Tidal Equations), whereas it can be computed much more simply from equilibrium tides if the ocean is deep enough.
Moreover, dissipation of ocean potential energy cannot be physically interpreted as heat dissipated per unit volume of the ocean, as is done in Figure 10 of Tyler's paper.
By contrast, it makes good sense to understand it as the integrated volumetric heating rate within the crust.
Finally, the ultimate location where the energy is converted to heat matters a lot because the amount of dissipation depends sensitively on the microscopic dissipation processes.
\item The ocean can be called the driver of tidal dissipation in two cases.
First, if the shell has a much smaller mass than the ocean and is soft enough that the tidal amplitude is nearly the one of an ice-free ocean (as on the snowball Earth and maybe Europa).
Second, if there are large dynamical effects in the ocean; but even so the elastic shell can have a strong controlling effect on the resonances by shifting them to much smaller depths.
For Enceladus, a plausible model is 20-25 km-thick shell above a 35-40 km-thick ocean,  much deeper than the resonance range.
The mass of the shell is thus comparable (in order of magnitude) to the mass of the ocean, and both are strongly coupled by the tidal potential.
Moreover, Enceladus's shell is quite rigid and strongly reduces the tidal amplitude of the ice-free ocean.
Mechanically, the most important role of a deep ocean is to provide a free-slip condition at the bottom of the shell, thus enhancing the tidal flexing of the shell.
If the ocean were always the driver of the shell, the fluid core of the Earth would also be the driver of mantle dissipation.
In fact, the concept of `dominant driving force' is not useful in a mechanically and gravitationally coupled system (mantle+ocean+shell).
The ultimate source of energy is the orbital and rotational energy of the interacting bodies.
\item It is misleading to say that the crust does not directly feel the tidal potential when it is modelled as a massless membrane.
In the membrane model, the gravitational response of the crust is separated from its elastic response: the former is modelled as a fluid layer added to the ocean, whereas the latter is modelled as a zero-thickness elastic membrane without mass.
If one is not happy with this approach,  the thick shell model of \citet{matsuyama2018} provides a treatment where the shell is massive and directly coupled to the tidal potential.
\item
Besides the cases of thin or stratified oceans considered in previous papers, Tyler proposes that dissipation of ocean tidal potential energy provides a plausible high dissipation state for Enceladus.
As discussed above, this kind of dissipation actually occurs within the crust and has been the focus in the last 40 years of many studies of icy satellites with subsurface oceans, e.g.\ \citet{cassen1980}, \citet{ross1987,ross1989}, \citet{ojakangas1989a}, \citet{hussmann2002},\citet{tobie2003,tobie2005}, and \citet{moore2006} (to cite only significant papers before 2006).
The convective model of the crust used by Tyler was initially tailored \citep{beuthe2016a} to illustrate strong damping of dynamical effects by crustal dissipation.
More realistic models of the crust predict much less crustal dissipation \citep{soucek2019,beuthe2019}.
As a caveat, I do not exclude the possibility that new dissipation processes could operate at the crust-ocean boundary, such as the tidally induced porous flow discussed by \citet{vance2007}.
\end{enumerate}
I will now give the technical details supporting my arguments.
In Section~\ref{EnergyBalance}, I rederive the energy balance equation on which Tyler's interpretation is based.
This is necessary because Tyler formulated his results in a very abstract way which is not easy to relate to the rest of the literature on ocean dissipation.
I also discuss the ambiguous meaning of the various terms in the energy balance.
These ambiguities can only be removed by computing each kind of dissipation (mantle, ocean, crust, or total) with another method.
In Section~\ref{DeepOceanLimit}, I show that dissipation of ocean potential energy tends in the deep-ocean/static limit to the total work of tidal forces.
In Section~\ref{CrustalDissipation}, I prove the equality between dissipation of ocean potential energy and crustal dissipation by computing the latter as the product of stress and strain rate within the crust.
A side result is the depth-dependence of the volumetric dissipation rate within the crust.
The notation for the Laplace Tidal Equations is defined in Appendix~\ref{AppendixA}.

\section{Energy balance}
\label{EnergyBalance}

Tyler computes dissipation by building an averaged energy balance equation from the equations of motion, as is common in oceanography (e.g.\ \citet{hendershott1972,egbert2001}).
Doing the same, I take the scalar product of the ocean momentum $\rho\mathbf{u}$ ($\rho$ being the ocean density) with the equation of tangential motion (Eq.~(\ref{LTE1})), use the vector identity ($\mathbf{u} \cdot \nabla f=\nabla \cdot( f \mathbf{u}) - f\nabla \cdot \mathbf{u}$), and substitute the equation of radial motion (Eq.~(\ref{LTE2})) in order to get
\begin{equation}
\partial_t {\cal E}_k
+ 2 \alpha \, {\cal E}_k
= - \mathbf{\nabla}  \cdot \mathbf{P}
- \frac{\rho}{D} \, {\rm Re} \bigg( \Big( g \sum_n\beta_n \, \eta_n - \upsilon_2 \, U_2^T \Big) e^{i\omega t} \bigg) \partial_t \eta \, ,
\label{LTEpower}
\end{equation}
where $\alpha$ is the coefficient of linear drag, $\eta$ is the radial tide (or radial displacement of the top of the ocean relative to its bottom, separated on average by the ocean depth $D$), while $(\beta_n,\upsilon_2)$ are complex parameters coupling the ocean to the crust and to the tidal potential $U_2^T$ (see Appendix~\ref{AppendixA}).
The rotation is synchronous so that $\omega$ represents both the mean motion and the angular velocity of rotation. 
In the LHS, ${\cal E}_k=\rho\mathbf{u} \cdot \mathbf{u}/2$ is the ocean kinetic power density while $2 \alpha{\cal E}_k$ represents the ocean dissipation due to linear drag.
In the RHS,  the horizontal power flux vector $\mathbf{P}$ is equal to
\begin{equation}
\mathbf{P} = \rho \, {\rm Re} \bigg( \Big( g \sum_n\beta_n \, \eta_n - \upsilon_2 \, U_2^T \Big) e^{i\omega t} \bigg) \mathbf{u} \, .
\end{equation}
Integrating Eq.~(\ref{LTEpower}) over the volume of the ocean $V_o$ and averaging over one orbital period $T=2\pi/\omega$, we see that the terms $\partial_t{\cal E}_k$ and $\nabla\cdot\mathbf{P}$ vanish.
The averaged energy balance equation can be written as
\begin{equation}
\dot E_O + \dot E_\eta = \dot E_W \, ,
\label{EnergyBalanceEquation}
\end{equation}
where
\begin{eqnarray}
\dot E_O &=& \frac{ \rho \alpha D}{T} \int_T dt \int_{S} \left( \mathbf{u} \cdot \mathbf{u} \right) dS \, ,
\label{EO}\\
\dot E_\eta  &=& \frac{\rho g \omega}{2} \sum_n {\rm Im}(\beta_n)  \int_{S} | \eta_n |^2 \, dS \, ,
\label{Eeta}\\
\dot E_W &=&  \frac{\rho \omega}{2} \int_{S} {\rm Im} \Big( \upsilon_2 \, U^T_2 \eta_2^* \Big) \, dS \, .
\label{EW}
\end{eqnarray}
where $S$ is the surface of the satellite.
Tyler interprets $\dot E_O$ as dissipation of ocean kinetic energy due to linear drag, which does not pose any problem.
Besides, he interprets $\dot E_W$ as the work performed on the ocean by the tidal forces.
This is correct as long as the mantle is non-deformable or both the mantle and the crust are elastic, as I will now show.

The mean rate of working by tidal forces throughout the body can be computed as a surface integral of product of the primary tidal potential $U^T$ and the time derivative of the secondary potential $U'$ induced by the deformations within the body \citep{zschau1978,platzman1984}.
When applied to the LTE coupled to a thin shell and a viscoelastic mantle, this formula yields the total dissipated power as a sum of three terms (Eq.~(H.4) of \citet{beuthe2016a}).
If the mantle is non-deformable, only the third term is non-zero so that
\begin{eqnarray}
\dot E_T =  \frac{\rho \omega}{2} \int_{S} {\rm Im} \Big( \upsilon_2^* \, U^T_2 \eta_2^* \Big) \, dS \, .
\label{ET}
\end{eqnarray}
Therefore, $\dot E_W$ coincides with the total tidal work as long as $\upsilon_2$ is real, which is for example true if the mantle is non-deformable and the shell is modelled as a membrane (Eq.~(\ref{CouplingConstants})).
But in the general case (thin viscoelastic shell with deformable mantle, or thick viscoelastic shell), $\dot E_W$ is not equal to the total tidal work $\dot E_T$.

Finally, Tyler interprets $\dot E_\eta$ as dissipation of ocean potential energy with the following justification: $\dot E_\eta$ is proportional to the potential energy of the ocean if the coupling constants $\beta_n$ are the same at all degrees (one should also assume that the mantle does not deform and neglect self-gravity).
He also interprets $\dot E_\eta/(SD)$ as the volumetric dissipation rate of ocean potential energy, which is an arbitrary choice because the term $|\eta_n|^2$ is not a depth-dependent variable.
Although Tyler considers dissipation of ocean potential energy as a generic mechanism, i.e.\ without specifying at first the origin of ${\rm Im}(\beta_n)$, the only example proposed in his paper is the coupling to a viscoelastic membrane (Eq.~(\ref{CouplingConstants})).
Since ${\rm Im}(\beta_n)=0$ if and only if the shell is elastic (Eq.~(\ref{iff})), Tyler's interpretation leads to a contradiction: ocean dissipation of potential energy is due to viscosity inside the shell.
Actually, the proportionality of $\dot E_\eta$ to the squared radial tide $|\eta_n |^2$ can be understood
in terms of the product of the stress and strain rate within the shell.
In Section~\ref{CrustalDissipation}, I will show  that $\dot E_\eta$ coincides with the crustal dissipation rate $\dot E_C$ computed from the microscopic dissipation rate in the solid shell, based on the standard approach of dissipation within solid layers.
As for the equality $\dot E_W=\dot E_T$,  the identification $\dot E_\eta=\dot E_C$ depends on the assumptions that the mantle is non-deformable and that the shell can be approximated as a membrane.

\section{Deep-ocean limit}
\label{DeepOceanLimit}

If the ocean is much deeper than the range in which ocean resonances occur, dynamical effects due to fluid motion become negligible and the degree-2 radial tide tends to the equilibrium (or static) tide:
\begin{equation}
\eta_2 \approx \frac{\upsilon_2}{\beta_2} \, \frac{U_2^T}{g} \, ,
\label{EquilibriumTide}
\end{equation}
while non-degree-2 components are negligible.
Crustal dissipation should also tend in that limit to the classical formula derived for static tides.
As in \citet{tyler2020}, I assume that the mantle is not deformable.
In that case, the degree-2 tidal Love numbers of the whole body (including the ocean and crust) can be computed in the static limit with the membrane approach (Eq.~(43) of \citet{beuthe2014}):
\begin{eqnarray}
k_2^T &=& \xi_2 \, h_2^T \, ,
\\
h_2^T &=& \frac{1}{1-\xi_2+\Lambda_2} \, ,
\end{eqnarray}
where $\xi_2$ is the degree-2 density ratio (defined after Eq.~(\ref{CouplingConstants})) and $\Lambda_2$ is the degree-2 membrane spring constant (Eq.~(\ref{MSC})).
Putting these relations together, I obtain
\begin{equation}
{\rm Im}(k^T_2) = - \xi_2 \, |h^T_2|^2 \, {\rm Im}(\Lambda_2) \, ,
\label{Imk2}
\end{equation}
while Eq.~(\ref{EquilibriumTide}) combined with Eq.~(\ref{CouplingConstants}) becomes
\begin{equation}
\eta_2 \approx h_2^T \, \frac{U_2^T}{g} \, .
\end{equation}
Substituting this equation into Eqs.~(\ref{Eeta})-(\ref{EW}), using Eq.~(\ref{Imk2}) and $g=(4\pi/3)GR\rho_{bulk}$ ($G$ is the gravitational constant, $R$ is the surface radius and $\rho_{bulk}$ is the bulk density), I can write that, in the deep-ocean limit,
\begin{equation}
\dot E_\eta \approx \dot E_W \approx - \frac{5}{2} \, \frac{\omega R}{G} \, {\rm Im}\big(k_2^T\big) \left( \frac{1}{4\pi R^2} \int_S |U_2^T|^2 \, dS \right) .
\end{equation}
For eccentricity tides, the term within brackets is equal to $(21/5)(\omega R)^4e^2$ (e.g.\ Table~1 of \citet{beuthe2013}, $e$ being the eccentricity), so that $\dot E_\eta$ tends to the classical formula for the total dissipation rate due to equilibrium or `static' eccentricity tides \citep{segatz1988}:
\begin{equation}
\dot E_\eta \approx \dot E_W \approx - \frac{21}{2} \, {\rm Im}\big(k_2^T\big) \, \frac{(\omega R)^5}{G} \, e^2 \, .
\end{equation}
Since this formula does not include fluid motion effects (at least if $k_2^T$ is computed in the static limit), it shows that $\dot E_\eta$ does not result from ocean dissipation.

\begin{figure}[h]
\begin{center}
     \includegraphics[width=12cm]{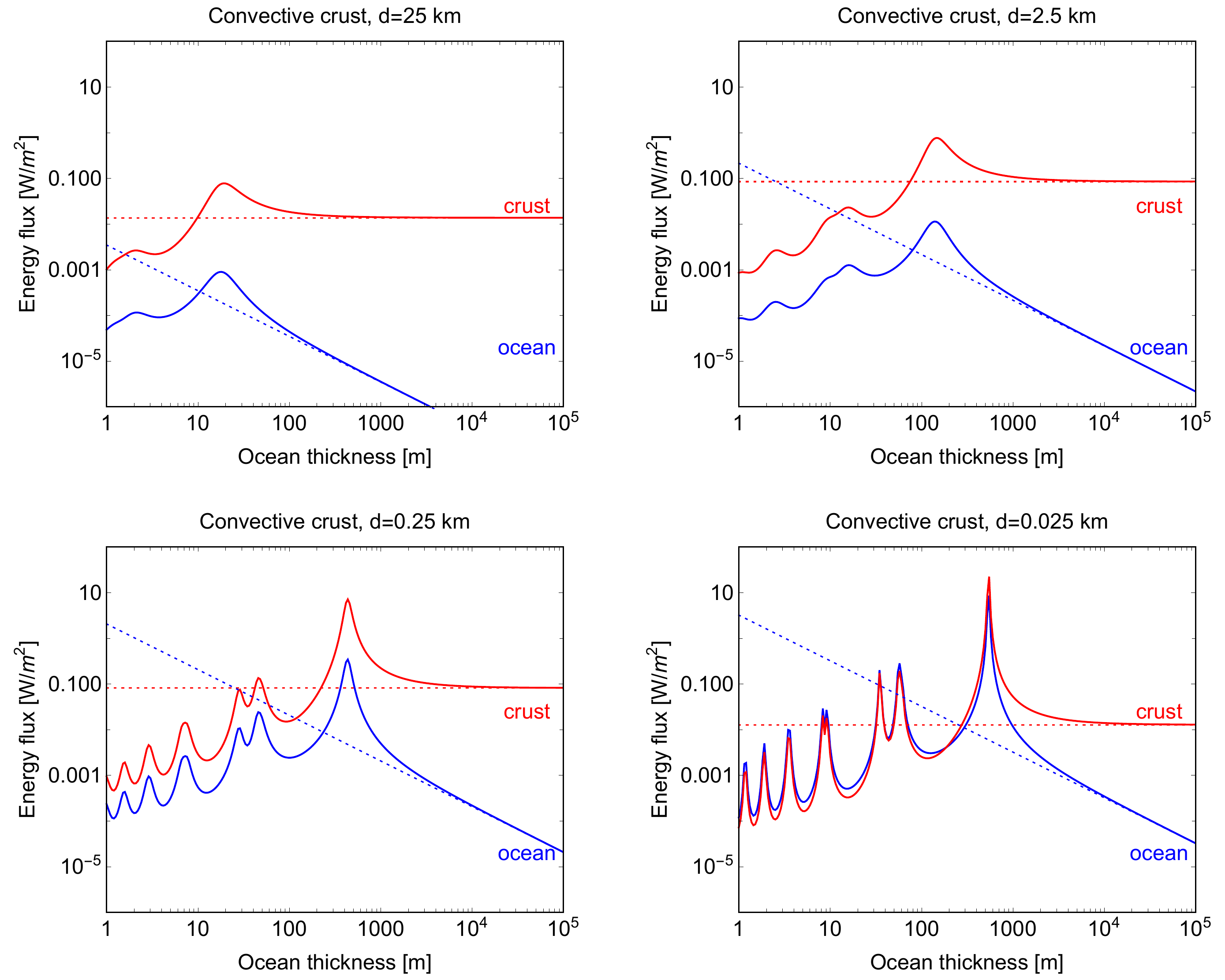}
\caption{\small
Tidal dissipation in the crust and ocean of Enceladus as a function of ocean thickness.
Each panel corresponds to a different value of the crustal thickness.
The forcing is due to eccentricity tides.
The crust is made of an upper conductive layer (40\% of crust thickness) and a lower convective layer (60\% of crust thickness).
Crustal rheology is modelled as in the convective model used in Figure 13 of \citet{beuthe2016a} (the asymptotes are described in the same paper).
}
\label{figVolumDissipRate}
\end{center}
\end{figure}

Figure~\ref{figVolumDissipRate} illustrates these conclusions by showing ocean and crustal dissipation (more precisely the corresponding surface fluxes) in Enceladus due to eccentricity tides.
In each panel associated with a different value of the shell thickness, the deep-ocean limit corresponds to the range of ocean thicknesses at the right-hand-side of the peaks.
Crustal dissipation tends to the constant value predicted from equilibrium tides, whereas ocean dissipation tends to zero.
The bottom right panel (very thin shell) is a case where the ocean is driving tidal dissipation in the resonance range.
The other panels show that a thicker shell shifts the resonance peaks to the left and strongly damps them.

\section{Crustal dissipation}
\label{CrustalDissipation}

In \citet{beuthe2016a}, I obtained the crustal dissipation rate by computing the average work done by the bottom load on the shell.
This procedure can give the wrong impression that this kind of dissipation occurs at the ocean-crust interface and is thus a new kind of ocean dissipation.
I will now show that the same result is obtained by integrating the microscopic dissipation rate over the volume of the crust.
For simplicity, I will assume that the material is incompressible, but the end result (Eqs.~(\ref{PC3})-(\ref{EC})) is the same if compressibility is included (see \citet{beuthe2014,beuthe2019} for similar computations including compressibility).

At each point within a viscoelastic solid, the dissipation rate due to friction is given by the rate at which stresses do work per unit volume, averaged over a tidal period \citep{kaula1963,kaula1964,peale1978}:
\begin{equation}
P(r,\theta,\varphi)
= \frac{1}{T} \int_0^T \sigma_{ij}(t) \, \partial_t {\epsilon}_{ij}(t) \, dt \, ,
\label{AveragedPower}
\end{equation}
where the summation convention applies.
Next, stress and strain are Fourier-transformed in the frequency domain (e.g.\ $\sigma_{ij}(t) = {\rm Re}(\sigma_{ij} \, e^{i \omega t })$) where the correspondence principle can be applied.
In an incompressible viscoelastic solid, stress is related to strain by $\sigma_{ij}=2\mu\epsilon_{ij}$, where $\mu$ is the complex shear modulus of the material which generally depends on position.
The volumetric dissipation rate becomes
\begin{equation}
P(r,\theta,\varphi) =  \omega \, {\rm Im}(\mu) \, \epsilon_{ij}^{} \, \epsilon_{ij}^{\,*} \, .
\label{PowerGeneralFormula}
\end{equation}
The application to a thin shell is best done in several steps.
First, the 3D strain in a thin shell can be computed in terms of two 2D functions: the stress function $F(\theta,\phi)$ and the radial displacement $w(\theta,\phi)=\eta^{top}$.
The dissipation rate expressed in terms of these variables is given by Eq.~(19) of \citet{beuthe2019}.
In the membrane limit, this equation greatly simplifies because the bending and mixed terms that depend on $w$ are negligible.
In that case, the volumetric dissipation rate within the crust reads
\begin{equation}
P_C(r,\theta,\varphi) =  \frac{\omega \, {\rm Im}(\mu)}{4\left(1+\nu\right) \mu_0^2 \, d^2} \left( \left| \Delta' F \right|^2 - (1+\nu) \, {\cal A}(F \,; F^*) \right) ,
\label{PC1}
\end{equation}
where $\nu=1/2$ is Poisson's ratio for an incompressible material.
The volumetric dissipation rate depends only on the radial coordinate $r$ through $\mu$ which can vary a lot from the surface to the bottom of a viscoelastic shell (it also varies laterally to a lesser extent).
By definition, the effective shear modulus $\mu_0$ ($\mu$ integrated over the shell thickness $d$) does not depend on depth.
The differential operators $\Delta'$ and ${\cal A}$ are defined in Appendices A and B of \citet{beuthe2018}.
Expanding the stress function in spherical harmonics ($F=\sum_n{}F_n$), I can write the action of these operators as (see Eq.~(B.11) of \citet{beuthe2018}):
\begin{eqnarray}
\Delta' F_n &=& \delta_n' \, F_n \, ,
\label{Op1} \\
{\cal A}(F_n \,; F_p^*) &=& - \frac{1}{4} \left( \Delta\Delta +\kappa_{np}' \, \Delta + \lambda_{np}' \right) (F_nF_p^*)  \, ,
\label{Op2}
\end{eqnarray}
where $\delta_n'=-(n-1)(n+2)$, $\kappa_{np}'=6-2(\delta_n'+\delta_p')$, and $\lambda_{np}'=(\delta_n'-\delta_p')^2-2 (\delta_n'+\delta_p')$.

\begin{figure}[h]
\begin{center}
     \includegraphics[width=6cm]{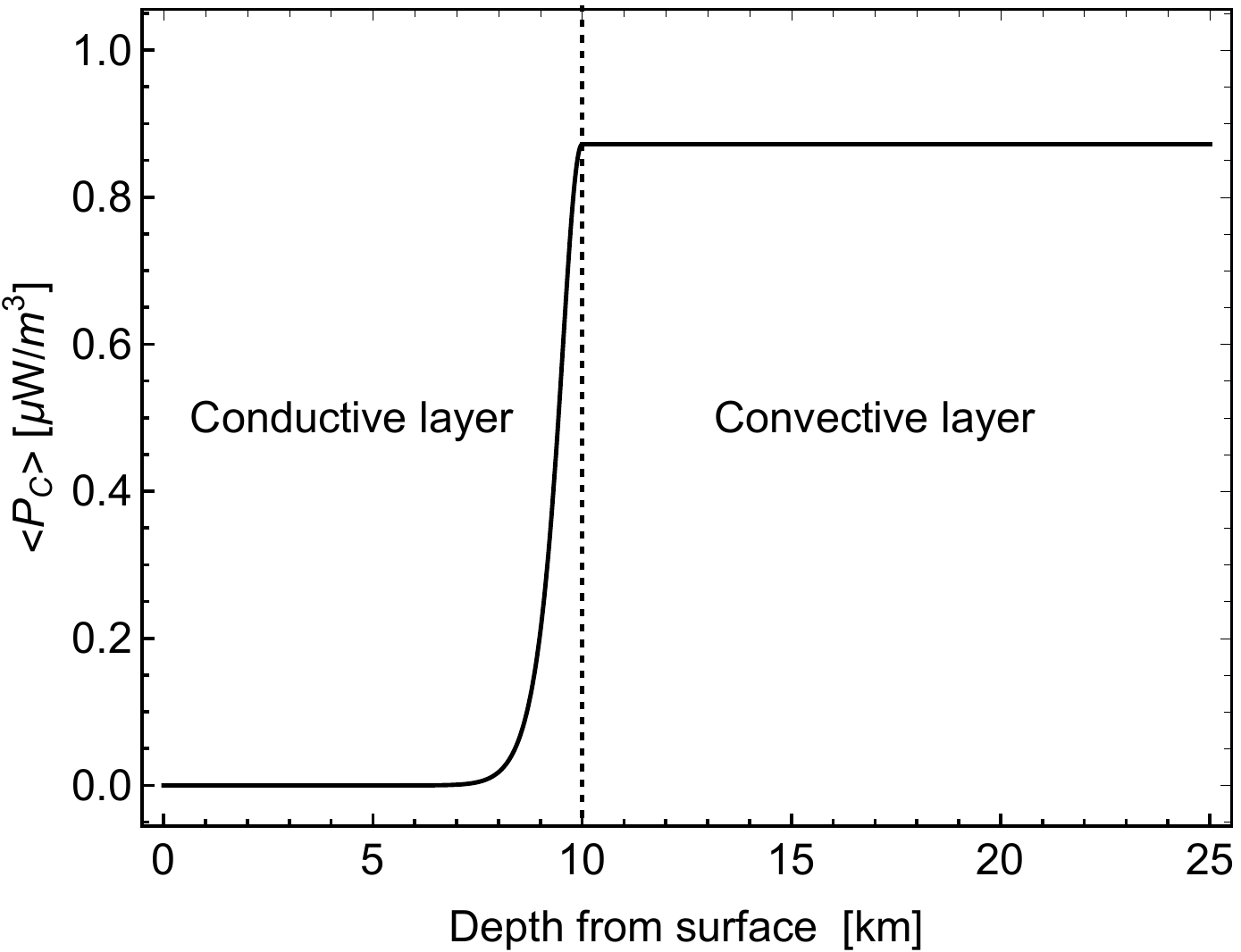}
\caption{\small
Tidal dissipation rate per unit volume of Enceladus's crust as a function of depth from the surface.
The forcing is due to eccentricity tides and the dissipation rate is averaged over the angles.
The ocean thickness is much larger than the resonance range.
The crust is made of an upper conductive layer ($10\,$km thick) and a lower convective layer ($15\,$km thick).
The rheology is modelled as in the convective model used for Figure 13 of \citet{beuthe2016a}.
}
\label{figVolumDissipRateDepth}
\end{center}
\end{figure}

Since we are not interested here by the spatial pattern of heating, I compute the angular average of the volumetric dissipation rate over the sphere $\Omega$ of radius $r$:
$\langle P_C \rangle=(4\pi)^{-1} \int_\Omega P_C \, \sin\theta \, d\theta \, d\phi$.
The orthogonality of spherical harmonics and the divergence theorem result in great simplifications:
\begin{equation}
 \langle P_C(r,\theta,\varphi) \rangle
= \frac{\omega \, {\rm Im}(\mu)}{16\pi \mu_0^2 \, d^2} \, \sum_n \frac{\delta_n'\left( \delta_n'-1-\nu\right)}{1+\nu} \int_\Omega  |F_n|^2 \, \sin\theta \, d\theta \, d\phi \, .
\label{PC2}
\end{equation}
If the shell is laterally uniform, the spherical harmonic component $F_n$ of the stress function is related to the spherical harmonic component $w_n$ of the radial displacement by Eq.~(59) of \citet{beuthe2018} (in which $\chi=\psi=1$ in the membrane limit):
\begin{equation}
F_{n} =  \frac{2\left(1+\nu\right)\mu_0}{\delta_n'-1-\nu} \, \frac{d}{R} \, \eta^{top}_{n}\, .
\label{Fn}
\end{equation}
Substituting Eq.~(\ref{Fn}) into Eq.~(\ref{PC2}) and using Eq.~(\ref{MSC}) for the membrane spring constant $\Lambda_n$ (recall that $\nu=1/2$), I can write the angular average of the volumetric dissipation rate as
\begin{equation}
 \langle P_C(r,\theta,\varphi) \rangle
= \frac{\rho g \omega}{8\pi d} \, \frac{{\rm Im}(\mu)}{{\rm Im}(\mu_0)} \, \sum_n {\rm Im}(\Lambda_n) \int_\Omega  |\eta^{top}_n|^2 \, \sin\theta \, d\theta \, d\phi \, .
\label{PC3}
\end{equation}
This formula does not change if compressibility is included, except that one should use the formula for $\Lambda_n$ depending on the effective Poisson's ratio (Eq.~(B.6) of \citet{beuthe2016a}).
Figure~\ref{figVolumDissipRateDepth} illustrates the depth-dependence of the volumetric dissipation rate inside the crust.
Dissipation is large in the convective layer and drops to zero in the conductive layer.
Finally, the crustal dissipation rate is obtained by integrating the averaged volumetric dissipation rate over the volume of the shell ($\dot E_C= \int_V \langle P_C \rangle \, dV$):
\begin{equation}
\dot E_C = \frac{\rho g \omega}{2} \sum_n {\rm Im}(\Lambda_n)  \int_{S} \big| \eta^{top}_n \big|^2 \, dS \, ,
\label{EC}
\end{equation}
where the surface integration is done at $r=R$.
This expression is identical to the work performed on the crust by the bottom load $q_n=\rho g\Lambda_n\eta_n^{top}$ (Eq.~(76) of \citet{beuthe2016a}).
It is also identical to $\dot E_\eta$ (Eq.~(\ref{Eeta})) if $\eta_n=\eta_n^{top}$ and ${\rm Im}(\beta_n)={\rm Im}(\Lambda_n)$, which is true if the mantle is not deformable (see Eq.~(\ref{CouplingConstants})).

\section*{Acknowledgments}
I thank an anonymous reviewer for carefully reading the manuscript.
This research has been supported by the Belgian PRODEX program managed by the European Space Agency in collaboration with the Belgian Federal Science Policy Office.

\begin{appendices}
\section{Laplace Tidal Equations for an ice-covered ocean}
\label{AppendixA}

Consider an icy satellite with a global subsurface ocean and submitted to tides synchronous with the rotation.
The dissipative Laplace Tidal Equations for the ocean sandwiched between a viscoelastic mantle and a viscoelastic shell (membrane, thin shell, or thick shell) are given by
\begin{eqnarray}
\partial_t \mathbf{u}
+ 2 \mathbf{\Omega \times u} + \alpha \, \mathbf{u} &=&
- \frac{1}{\rho} \mathbf{\nabla} p \, ,
 \label{LTE1} \\
\partial_t \eta
+ D \, \mathbf{\nabla \cdot u} &=& 0 \, .
 \label{LTE2}
\end{eqnarray}
The fluid motion variables are $\mathbf{u}(t,\theta,\phi)$, defined as the depth-averaged horizontal velocity vector, and the radial tide $\eta(t,\theta,\phi)=\eta^{top}-\eta^{bot}$, defined as the difference between the radial displacements at the top and the bottom of the ocean.
The term $\alpha\mathbf{u}$ describes dissipation due to linear drag with coefficient $\alpha$.
Other parameters are the body rotation vector $\mathbf{\Omega}$ with angular velocity $\omega$ (equal to the tidal frequency), the ocean depth $D$, and the ocean density $\rho$.
The operators $\nabla$ and $\nabla\cdot$ denote the surface gradient (tangential vector operator) and divergence (dimension $\sim m^{-1}$).
In the RHS, the forcing term is the surface gradient of a pressure-like potential $p$ given by
\begin{equation}
p = \rho \, {\rm Re} \bigg( \Big( g \sum_n\beta_n \, \eta_n - \upsilon_2 \, U_2^T \Big) e^{i\omega t} \bigg) \, ,
\end{equation}
where $g$ is the surface gravity.
The radial tide is Fourier-transformed in the frequency domain and decomposed over the sphere in spherical harmonics of degree $n$:
$\eta=\sum_n{\rm Re}(\eta_n e^{i\omega t})$.
The tidal forcing is represented in the same way, but only the harmonic degree 2 contributes:
$U^T={\rm Re}(U_2^T e^{i\omega t})$.
The complex parameters $(\beta_n,\upsilon_2)$ are defined by Eq.~(28) of \citet{beuthe2016a} if the shell is thin, and by Eq.~(22) of \citet{matsuyama2018} if the shell is thick.

The LTE have often been studied for an ice-free ocean above a non-deformable mantle, and without taking into account the self-gravity of the ocean, in which case $\beta_n=\upsilon_2=1$.
In general, both coupling constants differ from 1 and are complex if either the shell or the mantle is viscoelastic.
If the shell is thin and the mantle does not deform (as in Tyler's paper), $\upsilon_2=1$ but $\beta_n\neq1$ because of the ocean+shell self-gravity and the (visco)elastic response of the shell:
\begin{eqnarray}
\upsilon_2 = 1
\hspace{5mm} \mbox{and} \hspace{5mm} 
\beta_n = 1 - \xi_n + \Lambda_n \, ,
\label{CouplingConstants}
\end{eqnarray}
where the ocean+shell self-gravity depends on the degree-$n$ density ratio $\xi_n=(3/(2n+1))(\rho/\rho_{bulk})$.
The shell density is assumed to be equal to the ocean density, so that gravitation does not `see' any difference between shell and ocean, and  the mass of the shell can thus be included in the mass of the ocean.
The degree-$n$ spring constant $\Lambda_n$ can be computed given the rheology of ice and the shell thickness:
\begin{equation}
\Lambda_n = 6 \, \frac{(n-1)(n+2)}{2n^2+2n-1} \, \frac{\mu_0}{\rho gR} \, \frac{d}{R} \, ,
\label{MSC}
\end{equation}
where $R$ is the surface radius and $\mu_0$ is the effective shear modulus (the subscript stands for zeroth-moment), i.e.\ the complex shear modulus of ice integrated over the shell thickness $d$.
This formula is valid for an incompressible membrane, otherwise the spring constant also depends on the effective Poisson's ratio and on an additional bending term proportional to $d^3$ (Eqs.~(52)-(54) of \citet{beuthe2018}).
Eqs.~(\ref{CouplingConstants})-(\ref{MSC}) imply that
\begin{equation}
{\rm Im}(\beta_n) \neq 0
\hspace{5mm} \mbox{iff} \hspace{5mm}
{\rm Im}(\mu_0) \neq 0 \, .
\label{iff}
\end{equation}

\end{appendices}

\bibliographystyle{agufull04}
\renewcommand{\baselinestretch}{0.5}

\end{document}